\documentclass[sigconf]{acmart}

\usepackage{booktabs} 
\usepackage{balance}
\usepackage{filecontents}

\setcopyright{rightsretained}
\usepackage{tabularx}

\copyrightyear{2019} 
\acmYear{2019} 
\setcopyright{acmcopyright}
\acmConference[NSPW '19]{New Security Paradigms Workshop}{September 23--26, 2019}{San Carlos, Costa Rica}
\acmBooktitle{New Security Paradigms Workshop (NSPW '19), September 23--26, 2019, San Carlos, Costa Rica}
\acmPrice{15.00}
\acmDOI{10.1145/3368860.3368863}
\acmISBN{978-1-4503-7647-1/19/09}

\begin{document}
\balance
\title[Malware-based Manipulation of Web-based Email Discourse]{Manipulation of Perceived Politeness in a Web-based Email Discourse Through a Malicious Browser Extension}

\author{Filipo Sharevski}
\affiliation{%
  \institution{DePaul University}
  \streetaddress{243 S Wabash Ave}
  \city{Chicago}
  \state{IL}
  \postcode{60604}
}
\email{fsharevs@cdm.depaul.edu}

\author{Paige Treebridge}
\affiliation{%
  \institution{DePaul University}
  \streetaddress{243 S Wabash Ave}
  \city{Chicago}
  \state{IL}
  \postcode{60604}
}
\email{ptreebri@cdm.depaul.edu}

\author{Jessica Westbrook}
\affiliation{%
  \institution{DePaul University}
  \streetaddress{243 S Wabash Ave}
  \city{Chicago}
  \state{IL}
  \postcode{60604}
}
\email{jwestbro@cdm.depaul.edu}

\renewcommand{\shortauthors}{F. Sharevski, P. Treebridge J. Westbrook}

\begin{abstract}
This paper presents a specific man-in-the-middle exploit: Ambient Tactical Deception (ATD) in online communication, realized via a malicious web browser extension. Extensions manipulate web content in unobtrusive ways as ambient intermediaries of the overall browsing experience. In our previous work, we demonstrated that it is possible to employ tactical deception by making covert changes in the text content of a web page, regardless of the source. In this work, we investigated the application of ATD in a web-based email discourse where the objective is to manipulate the interpersonal perception without the knowledge of the involved parties. We focus on web-based email text because it is asynchronous and usually revised for clarity and politeness. Previous research has demonstrated that people's perception of politeness in online communication is based on three factors: the degree of imposition, the power of the receiver over the sender, and the social distance between them. We interviewed participants about their perception of these factors to establish the plausibility of ATD for email discourse. The results indicate that by covertly altering the politeness strategy in an email, it is possible for an ATD attacker to manipulate the receiver's perception on all of the politeness factors. Our findings support the Brown and Levinson's politeness theory and Walther's hyperpersonal model of email communication.  

\end{abstract}



\maketitle



\section{Introduction}

Email shapes much of our view of reality as people spend around three hours on average reading and replying to emails every day (depending on type of employment, social norms, practices etc.) \cite{Naragon}. An email discourse, especially in a formal setting, is asynchronous and offers little opportunity for shaping one's perception of interpersonal communication outside what can be pulled from the text. This allows senders to plan and revise messages before sending them, not just for grammar and mechanics, but also clarity and politeness \cite{Walther}. The theory of politeness, introduced by Brown and Levinson \cite{Brown}, suggests that people choose politeness strategies when formulating requests based on three factors: the degree of imposition in the discourse, the power of the receiver over the sender, and the social distance between them. 

The application of politeness theory in email discourse so far has been studied in a collaborative learning setting among students \cite{Vinagre}, \cite{Lam}, student requests to faculty \cite{Park}, \cite{Duthler}, \cite{Locher}, \cite{Bolkan}, and professional email negotiation \cite{Jensen}. The focus of these studies is concerned with the choice of a politeness strategy that the person making a request uses to shape their communication to the recipient in ways that will most likely prompt the recipient to carry out the request. However, none of the research discusses how the choice of the politeness could be manipulated by an automated software, residing "in-between" the requester and the receiver (and without their knowledge), to influence the perception of interpersonal communication.

We conducted a preliminary study to explore whether a change in the politeness strategy used to craft an email request, made through software and unbeknownst to the sender, affects the receiver's perception of the request on the three politeness factors. Based on the Brown and Levinson's politeness theory and Walther's hyperpersonal model of the email as a medium \cite{Walther}, \cite{Brown} our results show that it is plausible to manipulate the recipient's perception of the politeness factors in a web-based email discourse using a malicious browser extension. 
 
We developed the concept of ATD at the intersection of creative secure coding research and unfolding world events, particularly those involving cybersecurity and trustworthiness associated with the creating, sharing, disseminating, and filtering of online information \cite{Trowbridge}. A Washington Post article on Russian activity during the 2016 election describes the threat of information warfare in plain terms as: "Influence the information flow voters receive, and you'll eventually influence the government" \cite{Klaas}. Our preliminary research suggests that ATD is a novel approach to social engineering and information warfare that also has the potential to be used in a micro-targeted way to shape social reality \cite{Trowbridge}. The ATD attack is technologically feasible, and we detail a prototype that attempts to induce alternative sentiment and alter the social reality of a targeted user. An adversary interested in low-intensity information warfare might use the ATD approach beyond mere manipulation of social media ads or sending political spam. We believe that the preliminary results lay foundation for further research on countering ATD attacks in cyberspace. 


\section{ATD CONCEPTS}
 \subsection {Man-in-the-Middle Exploit}
An adversary can use malicious software to act as a man-in-the-middle in online communication, particularly in exchanging information through a web browser. Instead of merely "listening," to the data that flows between two people, they can induce misperception by changing text in a web browser through a malware-based extension. We call this new form of attack Ambient Tactical Deception (ATD). An adversary can change, remove or add specific words that change the email text without the knowledge of either the sender or receiver. Instead of defeating encryption or stealing credentials, an adversary might seek to influence the perception a targeted user has about another person by unobtrusively altering their online communication. 

The ATD attack is shown in Figure 1. In the first step, the adversary employs legitimacy-by-design (seeming legitimate both in visual design and in meeting what the user expects to see from a legitimate application) to persuade the target user to install a benign extension for a standard utility, for example a sticky notes extension like "Stickies" \cite{Vincent}, \cite{Newman}. This functionality is preferred because the extension requires text manipulation permissions from the user that later will be leveraged for the ATD attack. This will work because developing extensions for Chrome is free - a benign extension can pass all the security checks before publishing  \cite{Vincent}, ~\cite{Newman}.  

\begin{figure}[h]
  \centering
  \includegraphics[width=\linewidth]{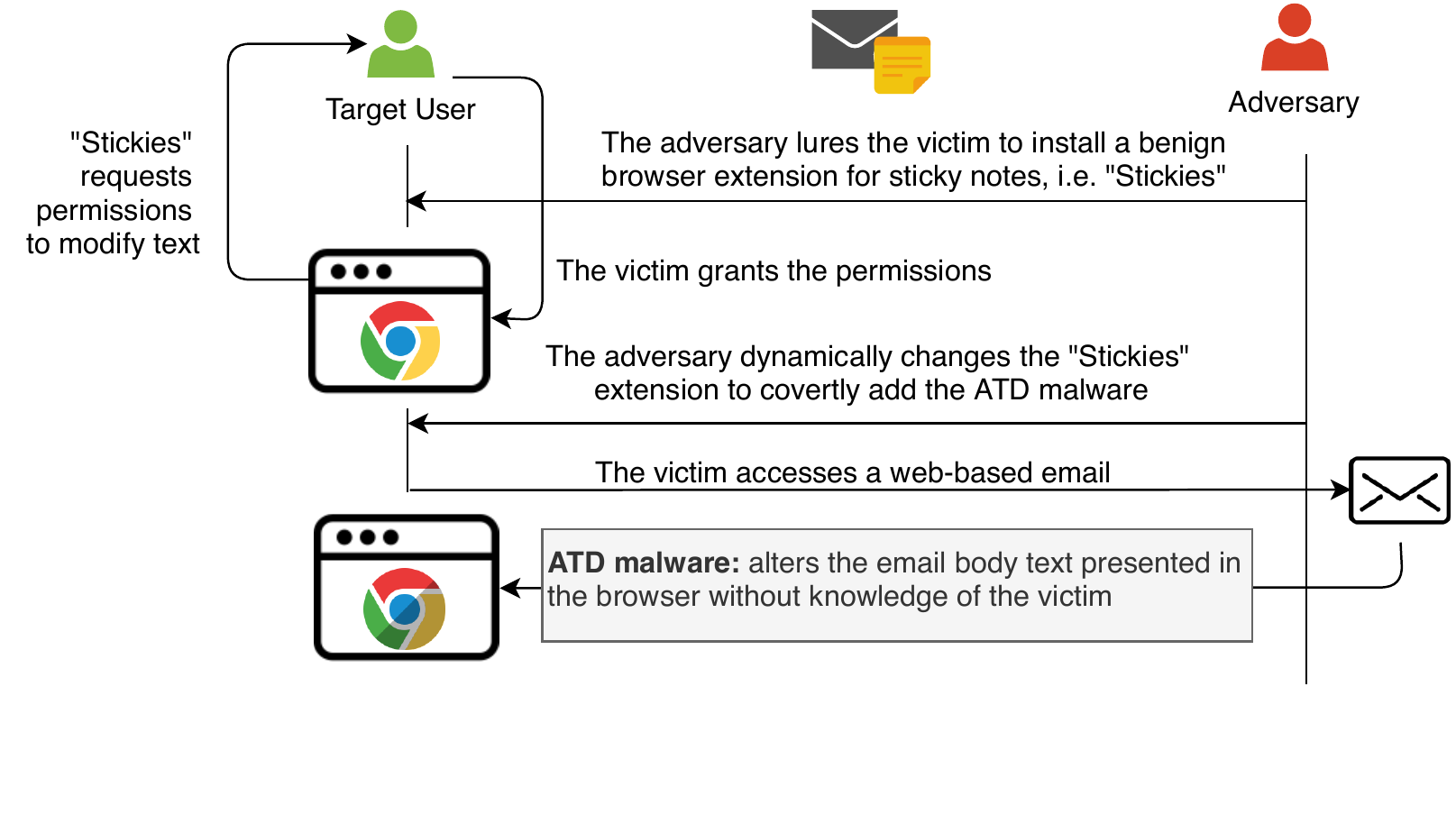}
  \caption{The ATD Attack Flowchart.}
\end{figure}

Assuming the targeted user is persuaded, the extension is installed on their system by asking permissions to change text, for example, to allow for copy/paste functionality. It is plausible that a user will proceed with installation when alerted for permissions. A study of Android users have found that 17\% paid attention to permissions during installation and only 3\% understand how permissions correspond to security risks \cite{Felt}. We assume these findings will generally hold true, given that Chrome and Android are part of the same Google ecosystem and users interact in a similar way with mobile applications and web browsers \cite{Bergman}.

The adversary can change the behavior of the extension dynamically after it is published and use previously issued permissions to manipulate any HTML text as part of the ATD attack. This change will go unnoticed because conventional web protections are focused on malicious text masked as code (injections) instead of seemingly bogus text \cite{Newman}. ATD also exploits the fact that Chrome is already a trusted application - when users give it permission to run certain code, like an extension, their operating system and most antivirus products give it a free pass. The altered text looks legitimate because it comes from a sender the user trusts as credible and reliable.

In our previous work we demonstrated an ATD proof-of-concept with a browser extension called "Sorry" \cite{Trowbridge}. The extension is intended to make "I" statements seem apologetic and craft an alternate reality for people who install the extension. An example application of "Sorry" used for an email request containing an "I" statement is shown in Figure 2 (a browser without the extension) and Figure 3 (a browser with the extension; the screen captures in both cases are taken at the same time). The legitimate email sent in Figure 2 is, unbeknownst to the sender and the targeted user, altered to look like the one shown in Figure 3. The targeted user has no reason to discard the email because he/she can verify the sender, the discourse topic, and the email request itself.

\begin{figure}[h]
  \centering
  \includegraphics[width=\linewidth]{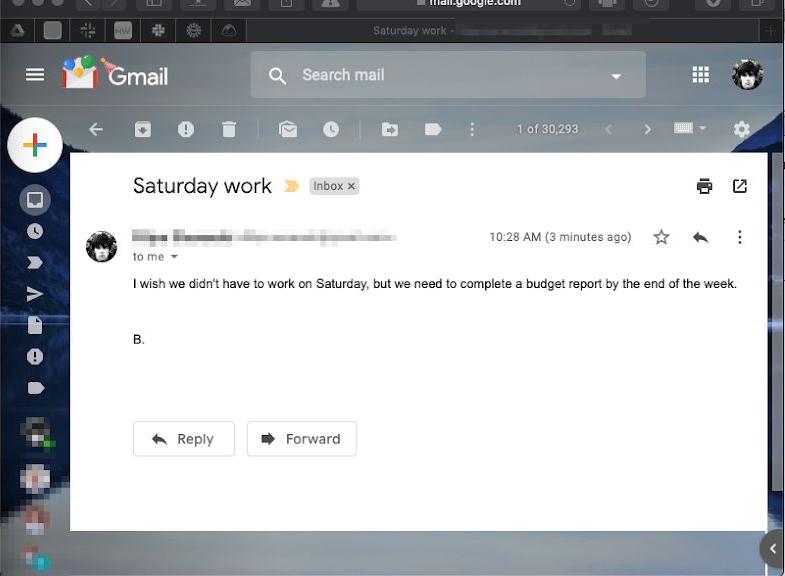}
  \caption{Email text in a browser \textit{without} the "Sorry" extension.}
\end{figure}

\begin{figure}[h]
  \centering
  \includegraphics[width=\linewidth]{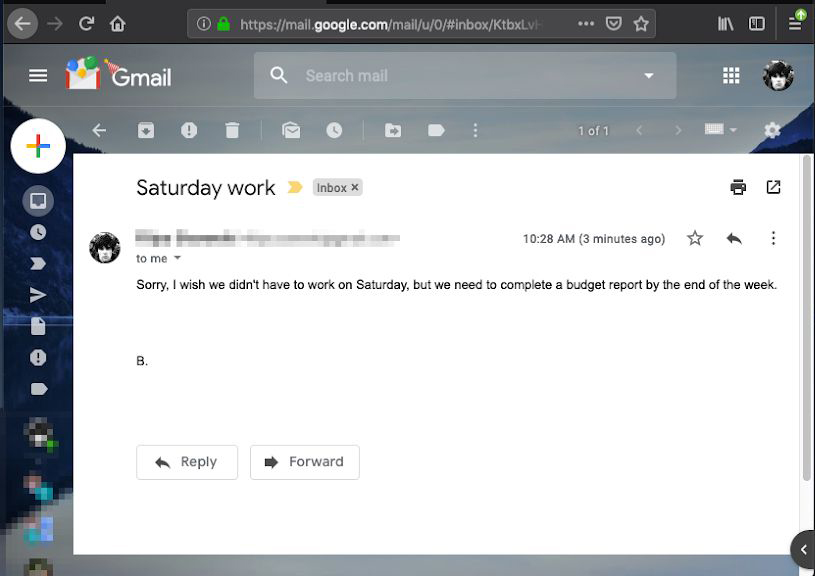}
  \caption{Email text in a browser \textit{with} the "Sorry" extension.}
\end{figure}

\subsection{Ambience}
Ambience in computing refers to "a digital environment that supports people in their daily lives in a nonintrusive way" \cite{Cook}. People expect that this ambience will "relieve the burden of daily chores, such as managing security, online correspondence, comfort, maintaining order, etc." \cite{Corno}. In a scenario in which a system is designed well, the computing ambience inevitably shapes people's perception of reality. When the goal is usability and the system becomes almost invisible, the user is able to complete tasks without considering the interface. With a goal to silently penetrate computing systems to manage perceptions, it's likely that adversaries will try to exploit the ambience. The tactical deception we are interested in exploring takes place via the ambient computing: the extension acts in the background as an unnoticeable intermediary for rendering web-based email text. To the degree that the changes remain unnoticed and manipulate only the wording but not the sender's address or the discourse context, they remain tactical and deceive the user.

\subsection{Tactical Deception}
The goal of deception is to "intentionally induce misperception" \cite{Spaf}. Inducing misperception in cyberspace is an attractive goal for several reasons. The primary mode of communication today is mostly computer-mediated, and deception can be focused on the messages between them \cite{Rowe}. Asynchronous messages, like email, are of particular interest because they avoid most of the nonverbal cues that reveal deceptions. Email allows more time for constructing a sound explanation, which in turn allows deception that withstands potential of scrutiny of the content ~\cite{Hancock}. 

Based on this, we explored the plausibility of a social engineering exploit in which a malicious, man-in-the-middle browser extension manipulates the content of emails to induce misperception. The tactic is to dynamically change the content of a web-based email accessed through a browser before it is displayed to a target user. This approach aligns with the objectives of cyber deception to negatively impact adversary activities or learn adversary tactics \cite{Urias}, \cite{Han} and psychological operations for "silent penetration of target's information and communications systems to manage perceptions and shape opinions" \cite{Cronin}. 

The term "tactical deception" is carefully chosen to delineate our work from the research on deception in Computer-Mediated Communications (CMC) \cite{Hauch}. Here too, the email is found to facilitate deception due to reduced nonverbal cues \cite{Toma}. However, the interest in CMC is on deceit that is deliberately created by the email sender. We are interested in deception as a tactic created by a malicious third-party, regardless of the email sender, that does not fabricate or introduce fake information into the communication. In an ATD attack, neither the sender nor the receiver are aware that their communication is slightly reorganized somewhere "in-the-middle."  

\subsection {Context and Purpose}
The advantage of ATD, from the perspective of an attacker, is that the social relationship between the target and another person, or people, can be manipulated without alerting any of the involved parties. ATD can be employed, for example, to alienate the target from other people. Our background research showed that it is much easier to make communication more negative than it is to make communication more positive \cite{Trowbridge}. This works to the advantage of an attacker who wants to make people unhappy with a person, a project, or a company. In a social engineering context, ATD could be used to slow down a competitor, poach disgruntled workers, or turn some parts of a group against a specific leader. 
 
In general, the ATD attacks do not need to be approached as a browser extension. Any situation in which a malicious actor has already achieved a man-in-the-middle advantage over a victim would be an opportunity for an ATD attack. We selected browser extensions as the example attack vector because they are a low-tech, minimal investment approach. Another attack vector for ATD might also be the LightNeuron malware which allows an attacker to read and modify any email passing through a compromised mail server \cite{Faou}. The ATD attacks are not necessarily focused only on email communication either. We have chosen email as an asynchronous type of computer-mediated communication, but ATD is equally applicable to the synchronous alternatives like WhatsApp, Line, and Facebook Messenger. Assuming a way to infiltrate into these communication platforms, an attacker could employ machine learning to make ATD less noticeable due to the synchronous type of interaction. Tone analyzers already exist and the attacker can learn the footprint of the personal communication between the target and another person or people \cite{IBMWatson}. Certainly, 
the change of communication will not always go unnoticed (for one, an honest typo in the text might trick the sentiment analysis algorithm, rendering the email/chat awkward to read, and making the ATD attack noticeable \cite{Grondahl}), but with machine learning there will be potential to scale and enhance the ATD attack. 

\subsection {ATD Paradigm}
We posit that ATD is a new social engineering paradigm - a model or pattern where textual (and possibly multimedia) content is manipulated in the context of online communication to distort one's mental picture or map of reality with the objective to establish psychological domination. Social engineering seeks to persuade people or gain a victim's compliance. ATD works by a silent, man-in-the-middle manipulation of content exchanged between two parties. The crucial difference in the ATD attacks is that the adversary is not "phishing" for a one-shot compliance but for a continuous compliance; the gain is not what the victim "has", (e.g. credentials, money, installation privileges, etc.) rather, the gain is what the victim "perceives." Unlike traditional phishing, ATD actually wants the victim to be engaged beyond a one-shot activity (e.g. clicking a link or opening an attachment). Unlike 419 scams, an ATD attacker makes no direct exchanges with a victim; instead they try to manipulate a legitimate multiple-shot activity between a victim and another communication party in a man-in-the-middle fashion. The catch with ATD is that this engagement must not cross the \textit{deception judgment threshold}, otherwise the victim will temporarily abandon the truth-default approach \cite{Levine}. The truth-default theory suggests that people presume others to be honest because they either don't think of deception as a possibility during communication or because there is insufficient evidence leaving them unable to prove they are being deceived.  

According to the Elaboration Likelihood Model (ELM), social engineering utilizes the "peripheral route" of persuasion to successfully engage the victim \cite{Cacioppo}. The ELM distinguishes "central" from "peripheral" routes of persuasion, where a central route encourages an elaborative analysis of a message's content, and a peripheral one is a form of persuasion that does not encourage elaboration of the message content (i.e. extensive cognitive analysis). Rather, it solicits acceptance of a message based on some adjunct element, such as perceived credibility, likeability, or attractiveness of the message sender, or "a catchy" phrase or slogan. Quite the opposite from the traditional social engineering, ATD utilizes the central route to encourage the victim to analyze the message's content and context (there is a possibility that other social engineering techniques might also utilize the central route of persuasion but ATD puts forward a means of automating this utilization). 

ATD attacks rely heavily on trust. Trust brings cognitive comfort that limits variety of thought and action and attention to detail \cite{Krishnan}. The ATD attack, unlike phishing, provides cognitive comfort to a degree that is necessary to "nudge" the victim to think about the context of a discourse or a request instead of the content \cite{Coventry}. For example, the ATD victims can verify the sender (email address, no attachments, mutual workflow, etc.) but might wonder whether they became distant with the sender if they suddenly address them with more polite emails. Of course, victims can be under-trusting and any such change might trigger the deception judgment threshold, in which case the ATD attacker needs to race to change their strategy before it is fully detected.  

However, some aspects from the obedience to authority theory can counter the potential of under-trusting victims uncovering the ATD attack. Obedience creates actions in deference to those who have perceived coercive power \cite{Weatherly}. The ATD itself doesn't create a perception of coercive power, but simply power of compliance. Victims, especially in workplaces, obey commands or requests to simply avoid a negative consequence such as disciplinary action (no one wants to "drop the ball"). As such, the ATD attacker can also play on the reactance of the victim, acting on a scarcity item such as time.

\subsection{ATD vs. Fake News}
Lazer et al. define fake news as "fabricated information that mimics news media content in form but not in organizational process or intent, i.e. lacking the news media's editorial norms and processes for ensuring the accuracy and credibility of information" \cite{Lazer}. Although fake news has been characteristic of social media platforms, email can also be a channel for spreading false information to purposely misinform people - not only in a political context, but also about topics such as vaccination, nutrition, and stock values. Unlike fake news, the ATD attack is not trying to mimic content but instead to reorganize it with the objective to manipulate perception on the receiver's end. In other words, the ATD attack acts as a silent, malicious editor with an intent different than the original sender of the email. Both fake news and ATD manipulate reality, but fake news alters facts whereas ATD changes perceptions. We have demonstrated that ATD is indeed effective in changing perceptions among Facebook users on the topic of freedom of political expression on college campuses \cite{NDSS}. 




\section{ATD OPERATIONAL ASPECTS}
\subsection {Threat Model}

The ATD attack we proposed and investigated is malware-based and requires only a small piece of software in a form of a browser extension that can be delivered to a targeted user via social engineering \textit{without} the sender and the receiver being aware of its existence. The malware provides an advantage in this case since the intention is manipulating perception, inducing alternative sentiment, and perhaps in the future, creating a completely alternative reality. The ATD attack, if deployed successfully, is independent of the email sender and the specificity of the email content. Moreover, the ATD threat is independent of any web page and works with sources that users usually trust and deem credible.

\subsection {Email Communication}
Email communication is asynchronous, limits inference, and allows senders to plan and revise messages for grammar, mechanics, clarity, and politeness \cite{Biesenbach}. This makes email a highly preferable vector for a malware-based ATD attack. Our initial focus is on politeness in emails because it is a critical component of human communication and personal discourse \cite{Park1}. Any manipulation of grammar, clarity and mechanics resembles a phishing email and might trigger what Levine calls a "deception judgement" by the targeted user: "If a trigger or set of triggers is sufficiently potent, a threshold is crossed, suspicion is generated, the truth-default is at least temporarily abandoned, the communication is scrutinized, and evidence is cognitively retrieved and/or sought to assess honesty-deceit" \cite{Levine}.

The theory of politeness, presented formally by Brown and Levinson \cite{Brown}, seeks to answer why people do not always speak in the clearest, most direct, and most efficient way possible. The reason, they suggest, is that we are all motivated by two desires: (1) the need for approval from or connection to others (positive face), and (2) the need to remain autonomous or independent (negative face) \cite{Duthler}. In order to maintain one's own positive/negative face, one must support the face needs of others. 

During the course of social interaction, conversational parties still need to make requests. A request intrinsically threatens another's face (positive or negative), so Brown and Levinson consider it as a Face Threatening Act (FTA). For example, simple requests threaten the receiver's negative face because the receiver's compliance with the request interferes with their desire to remain autonomous. Brown and Levinson propose that when confronted with the need to perform an FTA, the individual must choose between performing the FTA in the most direct and efficient manner or attempting to mitigate the effect of the FTA on the receiver's positive/negative face. The mitigation strategies are what they labeled "politeness strategies." Brown and Levinson propose four strategies along a continuum from least polite to most polite (requestive speech acts are added for illustration): 

\begin{itemize}
\item \textbf{Bald-on-record} - a way of speaking that is clear, direct, and concise. For example, "We need a budget, now!" 
\item \textbf{Positive politeness} - a redress directed to the recipient's positive face. For example, "Jake, we need a budget. Let's finalize it for the proposal today?"
\item \textbf{Negative politeness} - a redressive action addressed to the recipient's negative face, for example "Jake, I know you are busy, but would you be willing to meet with me for just an hour? We need a budget for the proposal - the deadline is today."
\item \textbf{Off-Record} - the receiver is given full autonomy to decide how to act upon the request. For example, a sender writing: "Proposals that include budgets are more likely to receive funding"  tries to implicitly note to the receiver that they need a budget to submit a complete proposal. 
\end{itemize}

The strategy an individual chooses to employ depends upon the \textit{weightiness} of the FTA (i.e. the extent to which the act is face threatening) \cite{Duthler}. The requester considers three variables when assessing weightiness:
 
 \begin{itemize}
\item The \textit{degree of imposition} associated with the FTA - a culturally and situationally defined ranking of impositions by the degree to which they are considered to interfere with one's desire for self-determination or approval (negative/positive face wants)
\item The \textit{power of the receiver over the requester} - the degree to which the receiver can impose their own plans and their own self-evaluation (face) at the expense of the sender's plans and self-evaluation
\item The \textit{social distance between the receiver and the requester} - a symmetric social dimension of similarity/difference within which the receiver and sender "stand for the purposes" of an act, and can refer to the frequency of interaction between them 
\end{itemize}

Holtgraves and Yang have found that increases in perceived power of the receiver, social distance, and the degree of imposition result in significant increases in the overall politeness of requests in interpersonal face-to-face communication. The requesters choose a more polite strategy either when the imposition is high, the receiver has a higher power, or the social distance is relatively high \cite{Holtgraves}. Morand and Ocker reviewed the politeness theory and stated that FTAs can be threatened in an email discourse just as it can in a face-to-face discourse \cite{Morand}. Holtgraves and Yang's findings are confirmed for email communication, too \cite{Park1}, \cite{Pemberton}. Using Walther's hyperpersonal model, Duthler proved that email facilitates manipulation of politeness (the choice of a politeness strategy) because senders have more time to craft a request to make the receiver create an idealized perception of their conversation partner \cite{Cronin}, ~\cite{Duthler}. 

\subsection {Targets}
The most likely targets of ambient tactical deception are "users that have abandoned traditional intermediaries," that include editorial judgement of the information provided \cite{Lin}. Any relationship in which people rely on web browsers for email correspondence would be good ATD targets. It is estimated that more than 44\% of the global email traffic is realized through a web browser (reading, writing, responding) \cite{SendGrid}. The targets must also be susceptible to social engineering to install the ATD extension in their browser in the first place (alternatively, the ATD extension can be preinstalled on the targeted user's machine without their knowledge). A study on susceptibility to online influence found that "individuals who demonstrate habitual e-mail use have increased susceptibility to social engineering." \cite{Vishwanath}

\section{ATD: Study}

\subsection{Motivation and Ethical Implications}
Based on Brown and Levinson's politeness theory and Walther's hyperpersonal model, our study was motivated by curiosity to see whether a change of the politeness strategy in an email request will change the receiver's perception of the politeness factors. ATD is a progression of social engineering beyond phishing and 419 scams, and in our view, the study is the first step to better understand how such manipulation might be prevented. ATD could be employed by an individual, but could be more effectively employed by an organization or government (as part of a psychological operations campaign, for example). The ethical implications are the same as those related to publishing any vulnerability: the value of publicly sharing the vulnerability with knowledgeable researchers outweighs the opportunity that potential attackers may benefit from the publication. If it is possible that the study introduces a viable attack, which we believe it will, we also believe that the attack would be discovered (and perhaps already has unbeknownst to us) by attackers. The study itself tests the plausibility of ATD with a proof-of-concept, locally-executed browser extension (not publicly available). In the context of a full-blown ATD, a responsible disclosure would entail contacting the implicated browser and mail server companies before making an extension or an email server exploit publicly available. 

\subsection{Background}
Politeness theory has been extensively tested in cross-cultural communication settings and suggests that the perceptions on imposition, power, and social distance vary between different cultures \cite{Blum}, \cite{Kulka}, \cite{Chen}, \cite{Economidou}. We focused on a discourse where both the sender and receiver have English as their first language because we crafted the politeness strategies in the email using the English lexical modifiers enumerated by Brown and Levinson (pp. 61-63) \cite{Brown}. Another reason is that the ATD malware-based extension we developed and used works only with English textual content. 

We chose to work with web-based email discourse in a formal setting for several reasons. A case study of politeness in the workplace found that Brown and Levinson's strategies have been largely employed in the Enron email corpus \cite{Peterson}. Further, a receiver in a formal setting is willing to carry out an email request. This is important to eliminate cases where a receiver discards a request as irrelevant. An analysis of the email responsiveness of the Enron corpus suggests that receivers are willing to carry out requests in formal settings and with a response generated within a short period of time \cite{Kalman}. Additionally, a receiver in a formal setting can easily verify a sender's email address. The social engineering research shows that the sender's email address verification is one of the main cues in deciding the legitimacy of an email \cite{Downs}, \cite{Sheng}, \cite{Gupta}.

We focused only on a scenario where there is no subordinate relationship between the sender and the receiver to avoid an a priori strong power assumption between the conversation parties (as is the case of student-faculty relationships \cite{Locher}, \cite{Bolkan}). Email is influenced by the context and hyperpersonal nature of communication and might not be suitable for any discourse. Previous research suggests that personal impressions from meetings in real life can influence how people perceive imposition, power, and social distance in emails \cite{Tidwell}. Therefore, we used only simple email requests focused on work tasks and outside of topics like politics, provocation, or humor \cite{Papacharissi}, \cite{Holmes1}, \cite{Holmes2}.

We also focused only on email discourse where the conversational parties have not met in person to avoid any outside influence on the perception of politeness. This is reasonable to expect with many workplaces allowing remote work and work in distributed teams. We instructed the participants to assume that they haven't met the sender, but that the sender is a work colleague. 

\subsection {Methodology}
A convenience sample of 36 participants (10 women and 26 men) from a mid-size midwestern university student research pool agreed to participate in the study. The study is qualitative in nature and we were more concerned with investigating the plausibility of ATD as a phenomenon rather than making a generalized hypothesis statement. For this purpose, the recommended sample size is between 30-50 participants \cite{Morse}. The inclusion criteria required participants to be 18 years old or above, to have at least one year of experience working in formal settings and communicating over email, and be a native English speaker (the ATD extension was developed to alter text written in English language and the theory of politeness pertains to western, English speaking cultures  \cite{Peterson}). The study was advertised as a "study in email effectiveness in a workplace" to prevent any influence the full knowledge about the ATD attack might have on the participants' response. The research involved minimal risk, was approved by the IRB, and the participants were debriefed on the overall study immediately after they provided their response. 

The participants were instructed to imagine a workplace scenario where the sender is a colleague from work and the participants are the receivers (We picked a gender-neutral name for the sender, "Sidney"). We also explained that the participants should assume that they can carry out the email request (e.g. they can deliver the budget, it is a part of their job description). Each participant was presented with a screen of the Chrome browser in a Windows operating system, and a Web Outlook client already opened in the browser. The email requests we used contained only grammatically correct textual content, without links, attachments, images, or emojis, so the participants were able to establish that the email content was legitimate. We crafted the emails to contain a neutral phrase, for example "We need a budget." This was important so the participants could recognize the literal meaning of the request in the first place \cite{Holtgraves}. 

The first email the participants read is shown in Figure 4 and employs the bald-on-record politeness strategy (the least polite). Once they were done, we asked them the following three questions: 

\begin{itemize}
\item What, in your opinion, is the degree of imposition in the email? 
\item What, in your opinion, is the power distance between the email sender and receiver? 
\item What, in your opinion, is the social distance between the email sender and receiver? 
\end{itemize}

\begin{figure}[h]
  \centering
  \includegraphics[width=\linewidth]{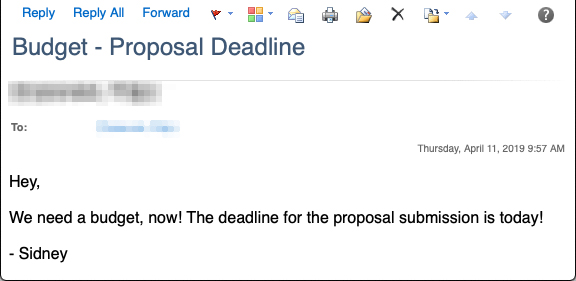}
  \caption{The first email with a \textit{bald-on-record} politeness strategy.}
\end{figure}

We repeated the same process with the second email, shown in Figure 5 where the email read as "I know you are busy, but would you be willing to meet with me for just a half an hour, we need a budget for the proposal - the deadline is today?" We chose the bald-on-record and the negative politeness strategy because they are the least and the most polite in a direct way (the off-the-record strategy is actually the most polite but it can introduce ambiguity given that it is left to the receiver to interpret the content; we wanted to avoid this). In a real ATD attack this could be done in an automatic manner for this and other email messages.


\begin{figure}[h]
  \centering
  \includegraphics[width=\linewidth]{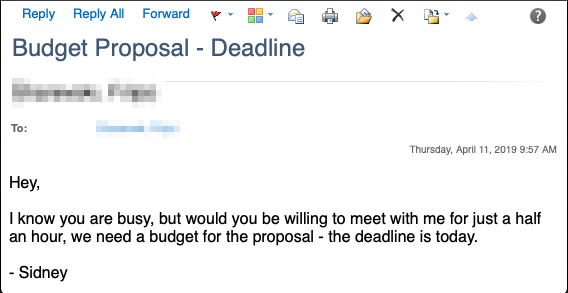}
  \caption{The second email with a \textit{negative} politeness strategy.}
\end{figure}

The participants provided verbal responses that were recorded, transcribed and coded for later analysis. Each participant was invited to the lab in person and given a 30 minute time slot. There was no overlap between participants. We conducted the de-briefing interview after the participants were done and asked them not to disclose any details about the study to other participants before they participated. 

\subsection {Research Design Considerations}
An important characteristic of our work is that ATD is employed only to alternate between politeness strategies and preserve the truthfulness of the email request itself; we are not engaging in what Sorlin calls a "manipulative discourse" - no manipulation of the FTA is employed, nor are we taking advantage of receivers' emotional vulnerability \cite{Sorlin}. We chose to use Brown and Levinson's politeness theory and Walther's hyperpersonal model for testing whether a perception can be influenced by automated and covert means through a dynamic manipulation of email textual content. We are aware of the complexity of interpersonal communication and various critiques to the theory and the model. 

The critiques on Brown and Levinson's politeness theory are directed towards the assumed sociolinguistic aspects of an interpersonal discourse \cite{Xie}, \cite{Eelen}, \cite{Goldsmith}. Brown and Levinson claim to provide a universally valid model of face that makes cultural norms irrelevant for the use of their politeness strategies \cite{Mao}. This claim, however, is debunked by Mao and Tarone who provided evidence that many cultures use politeness strategies differently \cite{Mao}, \cite{Tarone}. Along this lines, we limited our study only to western English speakers. ATD as a highly targeted attack; it thus needs to be tailored to the cultural norms of the discourse between the sender and the receiver to remain undetected. That requires modification of the malware-based extension, which is a complex task of its own, and subject for future research. 

Another critique of Brown and Levinson's politeness theory is that the four politeness strategies are not mutually exclusive \cite{Goldsmith}, \cite{Johnson}. We are well aware that emails can employ more than one politeness strategy. Clearly, the form of ATD attack we investigated won't work in this case. Therefore, we have limited the study to simple email requests with one politeness strategy employed. The politeness theory doesn't take into consideration external factors, for example mood or habitual interpersonal ways of interacting that can prompt individuals to choose a response in a specific discourse regardless of politeness strategies \cite{Goldsmith}. This certainly is a limitation that an attacker can't control and ATD will not work universally for all types of targets. Lastly, the politeness theory doesn't cover refusals of request in a discourse. To address this limitation we instructed the participants in our study to assume that the receiver is willing to carry out the email request. 

An alternative to Walther's hyperpersonal model, the "cues filtered out" model, suggests that email filters out all face-to-face cues, making a discourse impersonal and limiting the email as a potential avenue to employ any of the politeness strategies \cite{Jablin}. We acknowledge that the written and spoken politeness strategies differ due to a use of non-verbal cues. However, we believe that the hyperpersonal model of email communication is an appropriate vector for the ATD attack, especially in formal settings, because individuals regularly interact with emails (and rely on emails as a paper trail). The interaction and discourse at work entails face-to-face interaction, too, and this can certainly help targets detect the ATD attack. For this reason we instructed the participants to assume that they work on a remote team where email is their only method of formal communication. Many companies nowadays have remote teams and people working from home (or on the road), and we believe that an ATD attack can take advantage of this type of work structure. 

\section{Results}

\subsection {Degree of Imposition}
Thirty two out of 36 participants perceived an increase in the degree of imposition between the first email and the second email. For the first email, the perception was that the act is less face threatening because the sender was direct, clear, and concise without considering the face needs of the receiver. They felt that the email was not imposing because "it looks that the sender is definitely giving clear directions" or "trying to be cognizant of this person's time and feelings." For the second email, the perception was that the act became more face threatening to the receiver because the sender became more polite and considerate of receiver's time and plans. The second email looked like the sender is "more considerate of other's person time," "mincing words," "more polite and courteous," "acknowledging the effort of the receiver, sounds more like a meeting request rather a direct request."  

The reported change in perception of the degree of imposition from 89\% of the participants indicates that an ATD attacker can create a perceived context of formal domination over a victim. Not everybody in a formal setting becomes victim to a social engineering attack, however, and this holds true also for the ATD attack - four participants reported no change on how they are implied in the context of the task ("a budget has to be delivered by the end of the day in any case, regardless on how the email sounds"). Nonetheless, we believe that in the case of these four participants ATD was not noticed - the deception judgment threshold haven't been triggered to collect evidence to assess the truth-deceit of the email. Although this might not lead to an actual detection of the ATD attack, this strategy for workplace requests is sufficient enough to immunize targets at least from ATD aiming to interfere with one's self-determination to complete work tasks. 

\subsection {Power Distance}
Twenty-two participants stated that the receiver has more power over the sender in the second email. They felt that "now the roles are flipped," "the sender now knows the receiver is busy," and "the receiver can say, 'hey, I don't have time' and blow off the sender." Seven participants reported that the receiver's power increased in the second email and felt that the receiver now has equal power with the sender. Seven participants reported no change in the power distance between the sender and the receiver between the two emails: four of them reported the sender has more power, stating that "the (second) email is probably sent by a boss," while three reported the receiver has more power, stating that "the sender needs something from the receiver". 19.5\% of the participants that reported no change in the power distance further corroborate the possibility for targets to default to a simple conformity in a formal context of email communication. 


\subsection {Social Distance}
Twenty-five participants stated that the social distance between the sender is small in the first email and increased in the second email. They felt that the sender and receiver in the first email are "pretty casual," "seems they are in a pretty close relationship," but reported in the second email that the they "are not super cool with each other," "are more distant," and "more formal." Eight participants reported the opposite change in the social distance perception, stating that the sender and the receiver are much closer in the second email than in the first one. Only three participants, or 8.32\%, reported no change in the social distance between the sender and the receiver in both emails. The irrelevance of the social distance (no change in the perception) allows targets of ATD to immunize themselves from being engineered to psychologically attach to others by manipulating who they like and with whom they identify \cite{Allen}. These three outlier cases highlight an important characteristic of ATD: the attacker can exploit targets' continuance and affective commitment knowing that workplace friendships are found to improve organizational outcomes such as productivity and performance, as well as help individuals garner instrumental and emotional support \cite{Rumens}. This is the more disruptive aspect of the attack; to protect against ATD, employees have to revise their relationship footprints with colleagues at work to scrutinize for a potential ATD manipulation (at least in the beginning, it may very well be that a future AI system can detect and counteract ATD attacks).

\subsection {Summary}
The results are consistent with Brown and Levinson's politeness theory. The politeness strategy used in the first email was \textit{bald-on-record} (least polite) altered to a \textit{negative politeness} strategy in the second email (second to most polite). The politeness theory suggests that the choice of a more polite strategy is related to an increase in the degree of imposition, power distance, and social distance. The degree of imposition increased in the second email - 89\% of the participants stated that they perceived a bigger face threatening act in the second email. The perception of power between the first and second email increased in 80.5\%. The perception of social distance changed for 92\% of the participants between the two emails with half of them noticing a medium increase between the sender and the receiver. The relatively low percentage of users that didn't notice a change suggests that the ATD can go unnoticed (but not detected, which is a topic for a future study), but the perception of the email discourse for the majority of receivers will be covertly and successfully manipulated. This is consistent with Walther's hyperpersonal model of email and the possibility of editing the content for clarity and politeness. 

It is important to keep in mind that these findings may, or may not, illustrate that the the silent alternation of the politeness strategy is the sole factor contributing to the overall change of perception. It may well be that the ATD only acted as a behavioral intervention or a "nudge" in the cybersecurity behavior \cite{Coventry} - not to promote a best security practice, but rather to "socially engineer" a behavior to the objective of the ATD attacker. Other factors, including the study design, the content of the email, or the choice of politeness strategies, may have contributed to what participants reported about the politeness factors in both emails. We take this caution when analyzing the results and they should be seen as initial support of our idea to investigate the plausibility of ATD attacks rather than an authoritative test. 

\section{Discussion and Future Work} 
\subsection {Implications}
These results confirm what we previously proposed: that it is possible to use linguistic politeness as a vector for Ambient Tactical Deception (ATD) attacks. That we were able to achieve such a high rate of change in perception of imposition, power, and social distance is promising. As linguistic politeness is a fairly well-documented approach to communication, with definite rules, we feel that these early results could already be put into practice using procedural programming approaches and search/replace functions. Whether these approaches would directly affect the receiver of an email and their perception of the sender requires further research. Based on the current results, however, it is possible that simple ATD attacks that employ linguistic politeness algorithms would be far more effective than we were willing to speculate when we proposed ATD as a novel social engineering approach. 

The social engineering nature of ATD requires consideration of socio-economic factors and cultural norms, both for studying evolved forms of the attack and developing adequate defense. In our study we didn't focus on gender identity. In a more realistic scenario gender will certainly play a role in the choice of the politeness strategy, for example, Holmes suggested that women are more likely to use positive politeness than men \cite{Holmes}. Other factors such as age, race, nationality, status, sexual identity, or religion affect the relationship negotiability in a discourse and might complicate the assessment of weightiness when choosing a politeness strategy \cite{Kasper}. The demographic profile of the sender and the receiver factor in the ATD attacker's choice of target, email context, and politeness strategy. The \textit{ATD target profiling} is something we are currently working on. An important notion here is that the social attributes such as power and distance are themselves constituted by and subject to change in ongoing interaction. The directness is also subject to change depending on cultural norms, and every language has at its disposal a range of culture-specific routine formulae which carry politeness default values \cite{Ogiermann}. Unlike the static version of ATD we explored in this paper, the suggested dynamism and the culture-specific preferences in realizing requests have to be accounted for and realized accordingly in the malware for the ATD attack to be successful. 

\subsection {Technical Enhancements}
This current paper documents the potential of an ATD attack employing a linguistic politeness vector in a single email. A fully functioning ATD attack would require far more finesse to remain "ambient," i.e. invisible to the email recipient and the email sender. To prevent the email's sender from noticing that their words had been changed, an ATD extension would need to keep track of the changes it had made, and reverse those changes in any quoted text. It could become fairly complex to maintain ambience across an email thread, particularly if the thread involved more people. However, advanced ATD software systems could be developed to keep track of conversations, allowing for a bidirectional, or even multidirectional ATD attack: Bob sends an email to Alice, but it is intercepted by our ATD software. The email Alice receives is terse, and somewhat rude. Alice responds politely, but this advanced ATD software intercepts this response, and the response Bob receives is also terse, and very rude.

Employed at enterprise scale, an advanced ATD software could function across multiple accounts and devices, limited only by malicious actors' ability to gain and hold Man-in-the-Middle positions for each device or communication vector. With ATD that is focused only on text, this would include desktop, laptop, and tablet computers as well as smart phones. However, given the processing speed of contemporary computers and improvements in editing still images, video, and sound, as well as an increase in the amount of time people spend experiencing reality via some form of computer mediation, we can imagine a future in which an ATD attack changes the reality a victim perceives through "smart glasses" or "smart contacts." The ATD concept holds across any computer-mediated reality, and the ATD threat grows to the degree that people trust what they see, hear, feel, and perceive from any source that has passed through a computer of some sort. 

\subsection {Related Attacks}
The ATD attack intentionally induces misperception in an email discourse. An attacker can leverage the misperception of politeness to increase the likelihood of the receiver clicking on a link or opening a malicious attachment in the same or a follow-up email. The results indicate that the less polite an email is, the more the receiver perceives the sender as the "friendly and cool boss" \cite{percom}.  Under this cloak of a friendly and cool boss, the attacker can urge the receiver to take an action beyond simply reading the email. The peculiar incident with John Podesta, Hillary Clinton's campaign chef, where a Russian hacking group was able to retrieve a decade of his emails, comes to mind in this regard \cite{NYT}. He received a phishing email claiming that hackers had tried to infiltrate his Gmail account, and the sender provided a link to reset his password. Suspicious of potential phishing, he rightfully forwarded the email to the IT staff for further investigation. But then their reply contained a typo: it said the email was "legitimate" (instead of "illegitimate") so Podesta should proceed to change the password (and with that, reveal his new password to the hacking group). ATD is close to this incident in that the typo in the email could be made intentionally if the objective of the attacker is beyond manipulating perception but forcing a particular action from the targeted user. 

An interesting ATD attack in context of political fundraising is possible with the new tactics of targeted emails. A company called Grassroots Analytica provides campaigns with customized lists of donors who they believe are most likely to support them by sorting publicly available data spread across the internet \cite{Cohen}. An ATD attacker could use a similar algorithm and infer donors' (a) email addresses, and (b) the dollar amount these donors usually give (the targeted amount that most likely will be asked from them through a legitimate fundraising email). Our study confirms the politeness theory prediction that "the less polite the email, the least is asked from the receiver" - 32/36 participants perceived an increase in the degree of imposition between the less polite email and the more polite email. The ATD attacker can target generous donors with least polite emails to create a perception like "just give us the money," which can potentially make the donor give less or nothing.   

An attacker could also leverage the automatic conversion of emails written in a foreign language to English that Gmail offers for its users. Hypothetically, if the Soviet premier Nikita Khrushchev sent an email to the US ambassador saying "we will outlast you" (threatening with a nuclear arms race), then an ATD translation aiming to escalate the tensions between the US and Russia would have made the email read "we will bury you" (threatening with a nuclear attack) \cite{cia}. 

\subsection {ATD Detection and Prevention}
As an approach, ATD extensions only represent one way to employ ATD, and are a relatively low tech variant. Certainly, people who use multiple devices can detect inconsistency in the emails between their computer and smartphone (40\% of the people in a study on email adoption and usability reported they first open emails on their smart phone before they go to their web browser/email client \cite{Adestra}). People can also detect inconsistencies if they look into the email chain of the prior email discourse. ATD will have difficulties if people send only attachments in emails without text in an email discourse devoid of politeness or only reliant on attachments. Also, if people have been accustomed to seeing email politeness as insincere, or just a form of boilerplate, ATD targeting politeness might not work (e.g. bureaucratic emails).

If ATD goes unnoticed in these scenarios by the victims themselves, we would expect malicious software detection approaches to begin looking for web browser extensions that edit what is seen on the screen as the next line of defense on a browser-level. An example defense, also along these lines, would be using trusted browsers to detect JavaScript executions that are swapping words in the HTML text \cite{Kohlbrenner}. Another example is the Chrome's Manifest v3 API, which is designed to eliminate such extensions \cite{Google}. Content-level signing like email signatures, as implemented now, might not help in these regards because the ATD manipulation happens after the content integrity check happens in the sequence of email reception and display. 

Conceptually, ATD detection and prevention can benefit from application of the SCENE methodology for developing security nudges proposed in \cite{Coventry}. In the first step, \textit{scenario elicitation}; can be focused on the organizational (or interpersonal) way of crafting and responding to emails. Such an analysis will reveal the most commonly used politeness strategies in requests, in responses, and between different ranks. This input is next used in the second step for \textit{co-creating nudges} where a list of nudges is proposed (e.g. using organizational-specific keywords, wording, punctuation, possible use of emojis in the email, altering email between signatures, etc). A candidate nudge(s) are selected in the third step, prototyped in the fourth, and evaluated in the last step. 

An important consideration for detecting and defending from ATD is the scalability of the attacks. In our study we assumed a micro-targeted attack implemented as a web-browser extension that can also be realized as any type of specially crafted malware. Certain conditions have to be met for the attack to work, as discussed throughout the paper - a victim to be lured to install the web-browser extension (or an attacker to be able to compromise an email server) and the attacker to correctly infer the context of communication, for example the urgency with which a reader accepts email from various senders (or sincerity, for that matter - we all recoil from bureaucratic emails regardless of the politeness strategy employed \cite{Shuy}). A possible deterrent for ATD could be the diminishing return of micro-targeting a user to intentionally induce misperception on politeness - if an attacker has to spend immense effort on limited number of people then simply targeting a mass audience with fabricated email or social media content sounds much more effective. However, the Podesta incident suggests that the right ATD micro-targeting at the right time can be highly valuable to an attacker and have consequences beyond simply affecting the social relationship between the sender and the receiver. 

ATD, in essence, is an approach for psychological operations and information warfare as much as it is a social engineering exploit. In this context, a resourceful attacker might scale the attack instead by trying to create an entire browser, operating system (through system notifications), or email server that has the ATD functionality built in  - something that might sound improbable, on first thought, but nevertheless conceivable. Countries like China create local software, applications, and systems that certainly can be used to manipulate any text - online or local text (like version strings in an operating system) in an ATD fashion on a large scale, perhaps concealed under the premise of state-controlled censorship. This catalyzes a whole new set of implications worth exploring from a nation state or "fear, uncertainty, doubt" perspective. 

\section {Conclusion}
This paper explores the limited application of ATD via linguistic politeness using a web browser extension and does not represent the intense reality manipulation discussed in our first ATD paper \cite{Trowbridge}. We feel, however, that the difference is technical, rather than conceptual, and the issues with developing a fully-functioning ATD approach would be addressed by access to a larger team of developers, code fuzzers, and usability testers. These results are based on a small initial study. A more robust study would attempt to reproduce these results across multiple email scenarios, social media, and multiple linguistic politeness factors. The output of this study will be used to develop ATD detection and protection measures using the SCENE methodology and the trusted browser concept.

\bibliographystyle{ACM-Reference-Format}
\bibliography{sample-bibliography}

\end{document}